\documentclass{article}  
\usepackage{jamaica04}
\frompage{000} \topage{000}                                              
\def\rightmark{Exotic particle searches with STAR}
\def\leftmark{S. Kabana for the STAR Collaboration}
 
\title{Exotic particle searches with STAR at RHIC}

\authors{
{Sonia Kabana$^1$ for the STAR Collaboration
}\\[2.812mm]
{\normalsize
\hspace*{-8pt}$^1$ Yale University,\\ 
 06520 New Haven, USA  and
\\
\hspace*{-8pt} University of Bern,
\\
 3012 Bern, Switzerland 
}}

\abstract{
We present preliminary results of the STAR experiment at RHIC 
on exotic particle searches in minimum bias Au+Au collisions at
 $\sqrt{s_{NN} }$ = 200 GeV.
We observe a  narrow peak at 1734 $\pm$ 0.5 $\pm$ 5 MeV in the
$\Lambda K^0_s$ invariant mass  with
width consistent with the experimental resolution of about 6 MeV
within the errors.
The statistical significance
can be quantified between 3 and 6 $\sigma$ depending on cuts and methods.
If this peak corresponds to a real particle state it would be a
candidate for the $N^0$ or the $\Xi^0$ I=1/2 pentaquark states.
  }

 
\begin{document}
 
\maketitle
\setcounter{page}{1}

\section{Introduction}\label{intro}

\noindent
Hadrons made by 4 quarks and one antiquark, called pentaquarks,
were predicted 
long time ago (e.g. \cite{Preszalowicz}).
Pentaquark states have been searched for in the past without success 
until the recent finding of a candidate for the $\Theta^+$
 state first by the LEPS collaboration \cite{nakano} 
and by other experiments \cite{thetaplus}, motivated by
\cite{diakonov_polyakov_petrov_1997}.
Candidates for other pentaquarks have been 
 presented recently, in particular for the
 $\Xi^{--} (1862)$, $\Xi^- (1850)$, $\Xi^0 (1864)$
\cite{na49} and the $\Theta_c^0(3099)$ \cite{h1}.
\noindent
In this article we present preliminary  results 
of the STAR (Solenoidal Tracker At RHIC)
experiment at RHIC (Relativistic Heavy Ion Collider)
 on a search for the $\Xi^0$ I=1/2 
 as well as for the $N^0$ pentaquark states
in the decay mode $\Lambda K^0_s$.
 Since there may be many pentaquark multiplets
one may expect more than one $\Xi^0$ and $N^0$ state to appear.

\vspace*{-0.3cm}

\section{Experimental Setup}
\vspace*{-0.2cm}

\noindent
This analysis presents  results from data of minimum bias Au+Au collisions
at $\sqrt{s_{NN}} $=200 GeV, which were recorded in the 2001 run
with the STAR detector at RHIC.
A detailed description of the STAR experimental setup   can be found
in reference \cite{star}.
The present analysis is based on
 charged particle trajectories measured and identified with the help of a
large cylindrical time projection chamber (TPC) \cite{tpc} with full azimuthal
coverage, located inside a 0.5 Tesla solenoidal magnet allowing
for momentum reconstruction.
The TPC allows for the direct identification of charged particles with
small momenta 
by measuring their ionization
energy loss (dE/dx) \cite{tpc}.
Indirect identification of charged or neutral
particles decaying at least partly into charged measured
particles inside the TPC
 can be obtained with methods based on the topology of their
decay (e.g. $\Lambda$, $\Xi$, $\Omega$, $K^{\pm}$ etc decays) 
\cite{lambda_physrevlett89_2002_092301}.

\vspace*{-0.3cm}

\section{Data Analysis Techniques}

\vspace*{-0.2cm}

\begin{figure}[h]
\vspace*{-.1cm}
\begin{tabular}{ccc}
\hspace*{-0.2cm}
\hspace*{-0.2cm}
		\begin{minipage}{.31\linewidth}
\epsfysize 0.8\textwidth
\epsffile{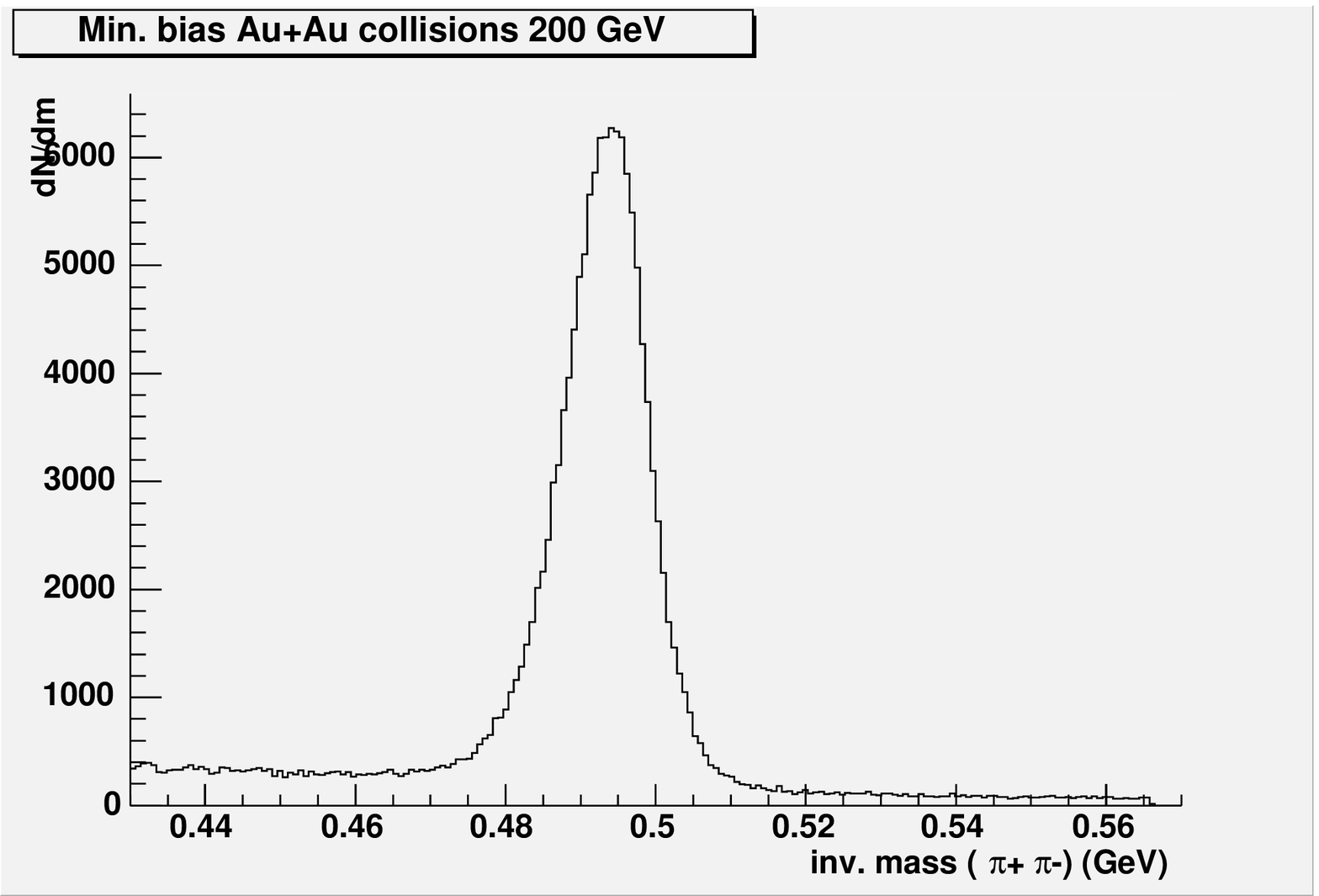}
\end{minipage} &
		\begin{minipage}{.31\linewidth}
\epsfysize 0.8\textwidth
\epsffile{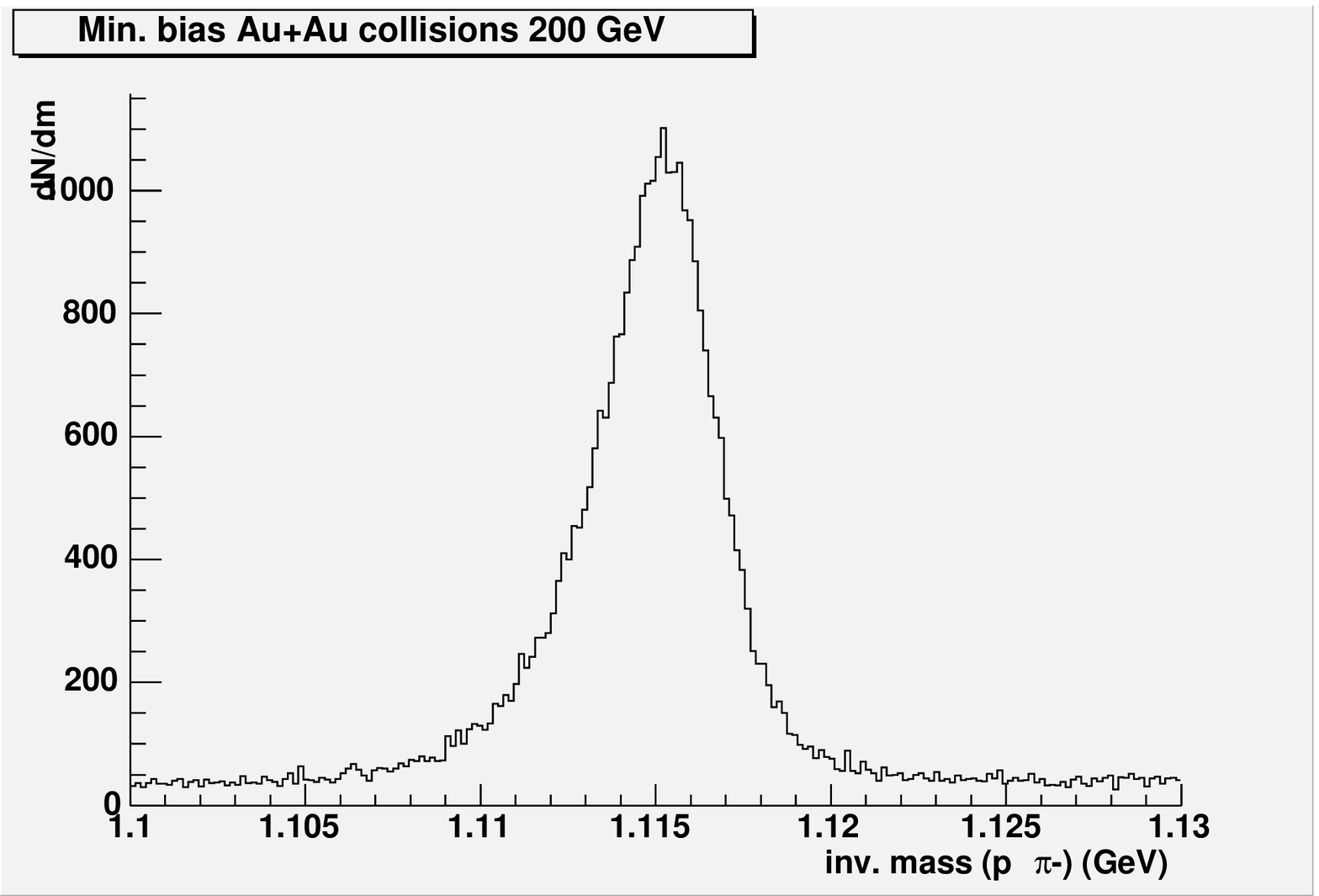}
\end{minipage}
\hspace*{0.2cm}
		\begin{minipage}{.31\linewidth}
\epsfysize 0.8\textwidth
\epsffile{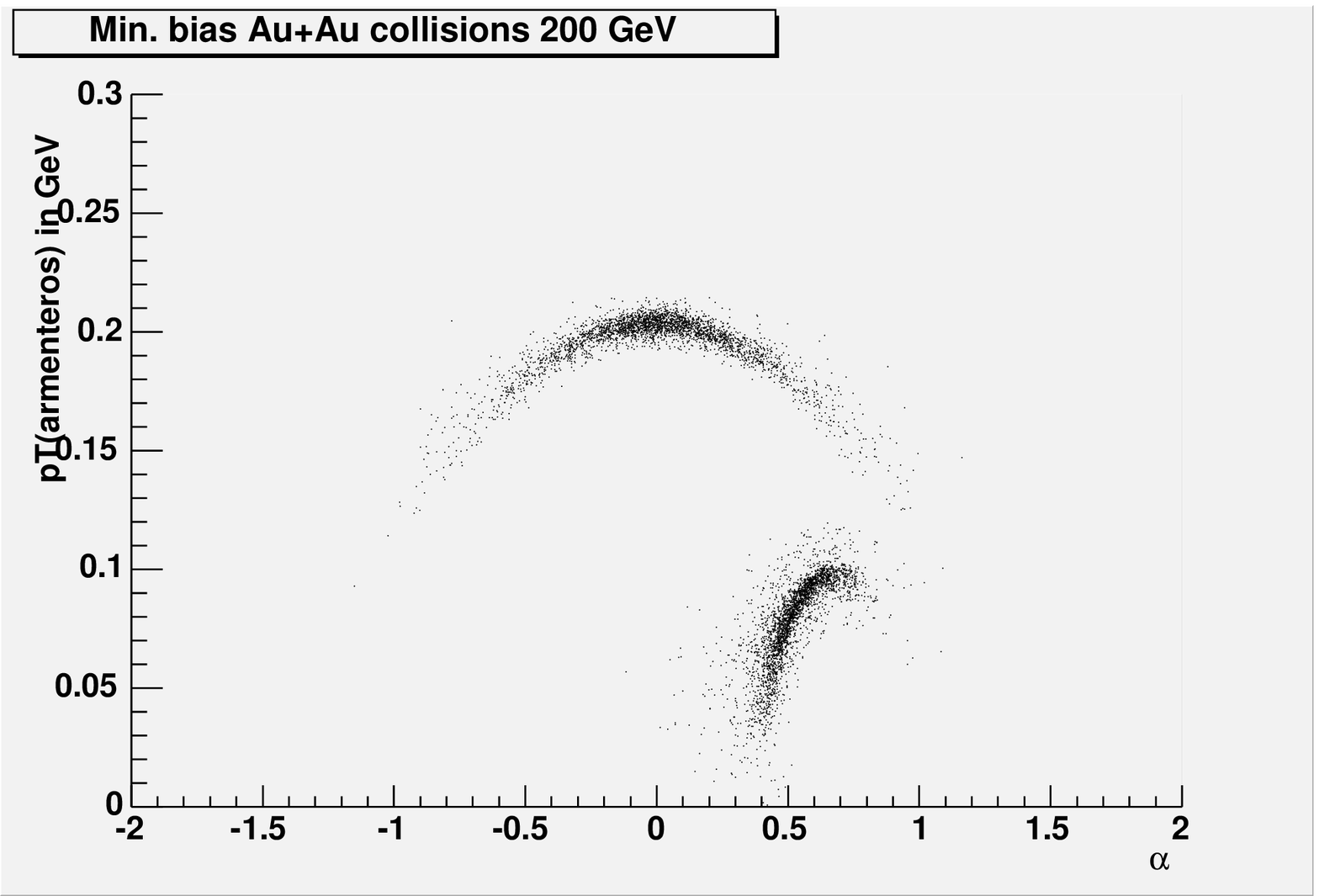}
\end{minipage}
\end{tabular}
\vspace*{-.4cm}
\caption{Figures left and middle: Invariant masses of $m(\pi^+ \pi^-)$ and
$m(p \pi^-)$ for the selected $K^0_s$ and $\Lambda$ candidates.
Figure right: Armenteros plot for the selected $\Lambda$
 and $K^0_s$ candidates.\label{arm}}
\vspace*{-.9cm}
\end{figure}

\noindent
In the present analysis we investigate the invariant mass of $\Lambda K^0_s$.
We required the Z position (Z is along the beam direction)
 of the main interaction vertex to be  $\pm$ 25 cm
around the center of the TPC.
The number of Au+Au events after this cut is 1.65 $ 10^6$
and comprise the total available statistics of minimum bias
Au+Au data at 200 GeV prior to the 2004 STAR run.
\noindent
We search for $V0$ topologies 
which are candidates for the decays $\Lambda \rightarrow p \pi^-$
 ($ \overline{ \Lambda } \rightarrow \overline{p} \pi^+$) and
$K^0_s \rightarrow \pi^+ \pi^-$
requiring the finding of a secondary $V0$ vertex, at least 6 cm away from the
primary interaction vertex.
The Distance of Closest Approach (DCA) between the positive and the negative 
$V0$ tracks (so called $V0$ daughters) is required to be less than 0.8 cm.
Each of the daughters is required to not originate from the primary
interaction  vertex (PV). In particular the DCA of the positive track to   
PV must be greater than 1 cm for the $\Lambda$ hypothesis,
2.5 cm for the $\overline{\Lambda}$ hypothesis and 1.3 cm for the
$K^0_s$ hypothesis.
The DCA of the negative track to
PV must be greater than 2.5 cm for the $\Lambda$ hypothesis,
1 cm for the $\overline{\Lambda}$ hypothesis and 1.3 cm for the
$K^0_s$ hypothesis.
Since the pentaquarks are expected to decay strongly 
we require the $V0$'s to point back to the primary vertex, in particular
to have a DCA to PV less than 0.4 cm.
\\
\noindent
We require at least 15 hits (out of maximally 45)
 for each track.
We require that the $\Lambda$ and $K^0_s$ candidates which are
combined to estimate their invariant mass do not share any tracks. 
The dE/dx of the $V0$ daughters
is required to
 be within 3 $\sigma$ around the expected dE/dx value for the
assumed particle hypothesis.
In order to enable the dE/dx identification 
we restrict the momenta of each  track in the region 
in which a good dE/dx identification is possible \cite{tpc},
namely below 0.7 GeV for (anti)protons and below 0.5 GeV for pions.
The latter cut on the momenta of the $\pi$ coming from the decay
of the $K^0_s$ is motivated independently
by Monte Carlo results for particles with mass in the range below
2 GeV and above the threshold for decay to $\Lambda K^0_s$,
which show that the momentum of the $K^0_s$ is
contained below 1-1.5 GeV, depending on the assumed initial
transverse momenta of the parent. 
We select the $K^0_s$ and $\Lambda$ candidates within a window of
 $\pm$ 35 MeV and $\pm$ 10 MeV
 around their mean invariant masses
which are shown in figure \ref{arm} (left and middle).
We accept only unambiguouisly identified $K^0_s$, $\Lambda$ 
and $\overline{\Lambda}$
 as illustrated in figure \ref{arm} (right) for $K^0_s$ and $\Lambda$.

\vspace*{-0.3cm}

\section{Results}
\vspace*{-0.2cm}

\begin{figure}[h]
\vspace*{-.3cm}
\begin{tabular}{c}
\hspace*{-.3cm}
		\begin{minipage}{.38\linewidth}
\epsfysize 1.79\textwidth
 \epsffile{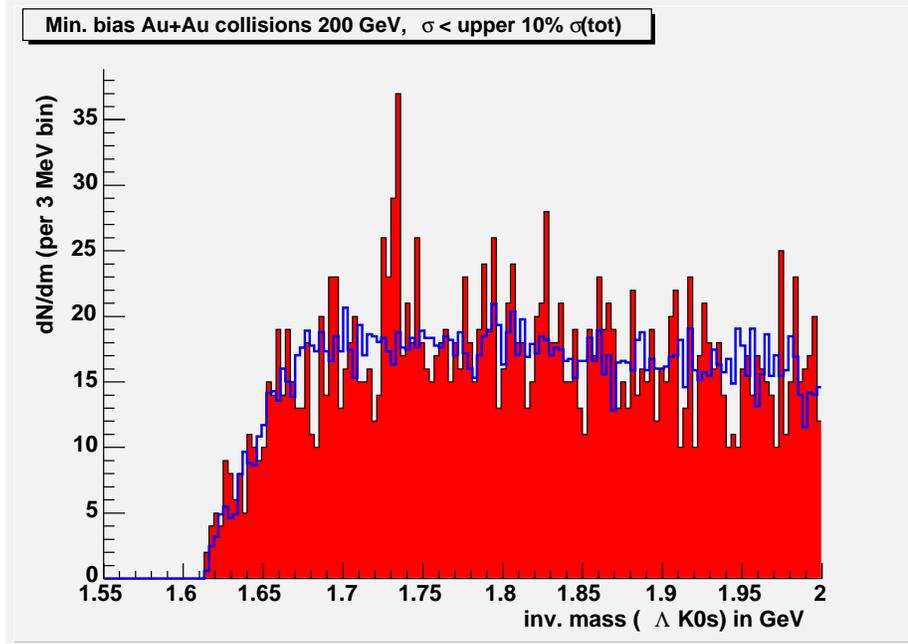}
\end{minipage} 
\end{tabular}
\vspace*{-.3cm}
\caption{Invariant mass distribution $\Lambda$ $K^0_s$ (in GeV) 
in min. bias Au+Au collisions
at $\sqrt{s_{NN} }$ = 200 GeV measured with the STAR experiment 
together with the estimated background using the mixed events technique
 (line).
We use 3 MeV bins.
The upper  $\sim$ 10\% of the $\sigma_{tot, Au+Au}$ has been excluded.  
\label{lk0s_1}
}
\vspace*{-.3cm}
\end{figure}

\noindent
Figure \ref{lk0s_1} shows the invariant mass of $\Lambda$ and $K^0_s$
which have been preselected as discussed in the previous section.
\noindent
The data are minimum bias Au+Au collisions at $\sqrt{s_{NN}}$ =200 GeV
while the upper $\sim$ 10\% of the $\sigma_{tot, Au+Au}$ 
has been excluded from the analysis to reduce the background
from the highest multiplicity events.
The  line shows the mixed event background expectation,
which has been calculated using $\Lambda$ and $K^0_s$ originating
from different events.

 \begin{figure}[h]
 \begin{tabular}{c}
\vspace*{-.4cm}
\hspace*{1.cm}
 		\begin{minipage}{.3\linewidth}
 \epsfysize 1.65\textwidth
 \epsffile{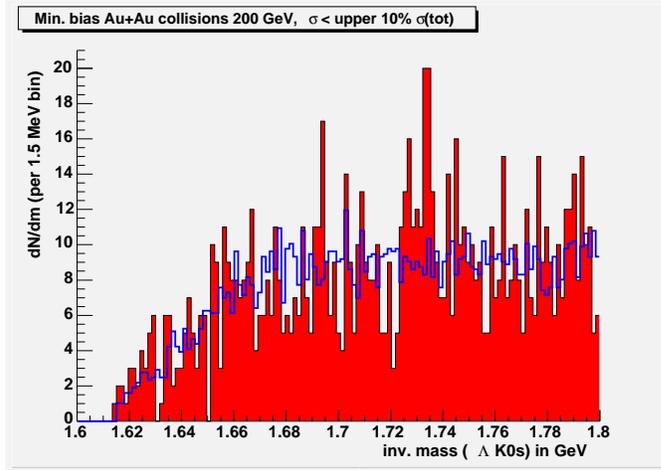}
 \end{minipage} 
 \end{tabular}
 \caption{Invariant mass distribution $\Lambda$ $K^0_s$ (in GeV) 
 in min. bias Au+Au collisions
 at $\sqrt{s_{NN} }$ = 200 GeV measured with the STAR experiment  
 together with the estimated background using the mixed events technique
  (line).
The upper  $\sim$ 10\% of the $\sigma_{tot, Au+Au}$ has been excluded.  
 We use 1.5 MeV bins.
 \label{lk0s_1_1}
 }
\vspace*{-0.3cm}
 \end{figure}

\noindent
The mixed event distribution has been normalized to the 'signal+background'
 distribution in the region below 1.7 GeV and between 1.76 and 2 GeV.
 Other choices of smaller normalization regions did not influence the
 results significantly. 
\noindent
We use bins of 3 MeV.
We observe a  peak at 1734 MeV.
When we fit this peak with a Gaussian distribution restricted to the
region of $\pm$ 3 MeV 
around the mean, plus a polynomial
function for the background from 1.65 to 1.8 MeV,
using 1 MeV bins,
we obtain  a gaussian width of
4.6 $\pm$ 2.4 MeV  and a
 $\chi^2/DOF= 1.09$.

\noindent
We obtain similar results 
when using a Breit Wigner distribution
and different bin sizes.
The width is consistent with the
experimental resolution  within the errors.
 The latter has
 been estimated with Monte Carlo
generated particles with mass 1730 MeV, flat $dn/dy$ distribution
$\pm$ 1.5 units around midrapidity, exponential spectrum in 
$m_T$ with inverse slope 250 MeV  and Breit Wigner width
of 1 MeV, decaying into $\Lambda K^0_s$,
 which have been tracked through the Geant STAR simulation
 and have been embedded in real p+p STAR data.
We could reconstruct the initial Monte Carlo particles with a
mass of 1729 $\pm$ 0.7 MeV and a width of 6.3 $\pm$ 1.7 MeV
which is dominated by the experimental resolution.

 \begin{figure}[h]
 \begin{tabular}{c}
\vspace*{-.3cm}
\hspace*{1.4cm}
 		\begin{minipage}{.3\linewidth}
 \epsfysize 1.65\textwidth
 \epsffile{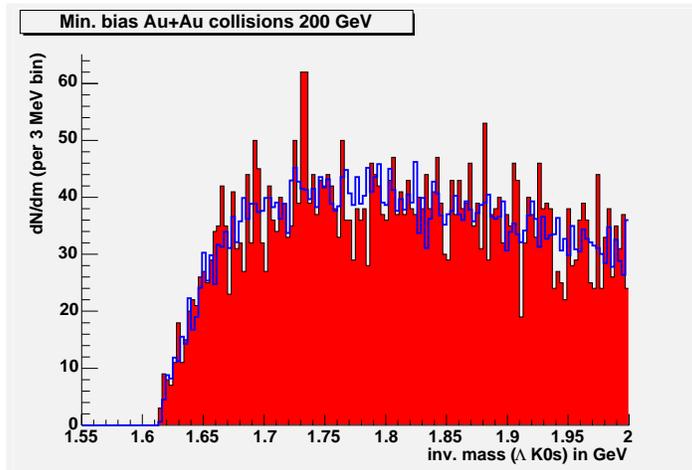}
 \end{minipage} 
 \end{tabular}
 \caption{Invariant mass distribution $\Lambda$ $K^0_s$ (in GeV) 
 in min. bias Au+Au collisions
 at $\sqrt{s_{NN} }$ = 200 GeV measured with the STAR experiment (red)
 together with the estimated background using the mixed events technique
  (line).
 We use 3 MeV bins.
 \label{lk0s_2}
 }
\vspace*{-0.7cm}
 \end{figure}

\noindent
The gaussian fit to the data described above
 results in a mass of 1733.6 $\pm$ 0.5 MeV with a systematic
error of $\sim$ 5 MeV.
The latter has been deduced from the deviation of the $K^0_s$ mass from
the mean $K^0_s$ mass given by the PDG \cite{pdg} in the 
appropriate region of transverse momentum.
\noindent
The values for the mean position of the peak and the width
obtained by using Breit Wigner fits and/or Gauss fits
  and/or different binning and/or different
analysis cuts were consistent with the above values within the
statistical errors.
Assuming the mass of 1734 MeV and considering
the region of $ \pm$ 1.5 $\sigma$ around the mean of the mass
we obtain a significance of 
 $S/ \sqrt{ B} $ = $30.6/ \sqrt{ 35.4}$ = 5.1,  while a more
 conservative estimate gives
 $S/ \sqrt{S+B} $ = 3.8. 
The ratio of the signal  
to its own error is $S/\sigma(S) = S/ \sqrt{ S+2B} $ = 3.04. 
Furthermore we observe a peak at 1693 $\pm$ 0.5 (stat) $\pm$ 5 (syst) MeV
with a significance of $S/ \sqrt {B}$ = 2.92. 
The width is similar to the previously discussed peak at 1734 MeV.
\\
 Fig. \ref{lk0s_1_1}  shows the same data as fig. \ref{lk0s_1}
however with bin size 1.5 MeV, in order to illustrate
that the 1734 peak in fig. \ref{lk0s_1} is not assymetric,
but the left shoulder seen there
 is a well separated peak at mass 1726.6 $\pm$ 0.5  MeV.
\\
Figure \ref{lk0s_2} shows the same distribution as figure
 \ref{lk0s_1} but without any
restriction on the event centrality.
The significance  is 
 $S/ \sqrt{ B} $ = $40.55/ \sqrt{ 83.45 }$ = 4.44 in this case.
We obtain the best significance of $S/ \sqrt{ B} $ = $19.4/ \sqrt{ 10.6}$= 5.93
for semiperipheral events.

\vspace*{0.2cm}

\noindent
In the following we discuss some possible sources of systematic errors.
One source of systematic errors especially in high multiplicity events
is the fact that the tracking software may split a real trajectory into
two tracks which will have a very similar momentum
and can lead to selfcorrelations and to
peaks in invariant mass distributions.
We excluded split tracks from the analysis
by demanding at least 25 hits  on each track, as the maximum
number of hits is 45.
We also studied the effect on the invariant mass $\Lambda K^0_s$ 
if one selects either the same track taken e.g. as pion of the
$\Lambda$ and of the $K^0_s$ or if one selects only the (same charge) tracks
with  momentum differences  below (respectively above) e.g. 100 MeV.
The 100 MeV is well above the expected momentum difference on the basis of the
momentum resolution in this momentum range \cite{tpc}.
None of these studies gave a peak near 1734 or 1693 MeV.
\\
\noindent
Another possible source of systematic errors
is the possibility that the proton of the $\Lambda$ and the $\pi^-$
of the $K^0_s$ or the $\Lambda$ come from a $\Delta(1232)$ decay.
As the sum of the  $\Delta(1232)$ mass and the $K^0_s(497.7)$ mass
gives 1729.7 MeV this could lead to a peak at 1734 MeV,
given  the STAR systematic error of about 5 MeV.
This possibility has been investigated either by cuting harder on the DCA
of all tracks to the PV (as the $\Delta$ decays right at the primary vertex)
or by cuting out cases giving an invariant mass for 
 $p( \Lambda) + \pi^-(K^0_s) $ outside (respectively inside)
 a window of $\pm$ 30 MeV around the $\Delta(1232)$ mass.
These studies showed that the peak at 1734 MeV
cannot be understood as due to the $ \Delta(1232)$+$K^0_s(497.7)$
mass reflection.

\vspace*{-0.3cm}

\section{Discussion}\label{discussion}

\vspace*{-0.2cm}

\noindent
In this section we discuss the possible assignment of the
observed peaks, if they would correspond to 
real particle states. 
The already known particles decaying into $\Lambda K^0_s$
with mass near 1734 MeV namely the N(1710) and N(1720)
 (with Breit Wigner masses in the range up to 1740 MeV)
have a width of 100 MeV or greater and
are therefore not good     
                                                 candidates for the narrow peak seen here at 1734 MeV.
This peak is not a candidate for the decay of the $\Xi^0$ pentaquark
with isospin 3/2, as the latter does not decay into $\Lambda K^0_s$ due to
isospin violation.
It is a possible candidate for two pentaquark states: the 
$N^0$ with quark content 
$udsd \overline{s}$ 
 decaying into $\Lambda K^0 $
and the $\Xi^0$ with isospin I=1/2 and quark content
$udss \overline{d}$ decaying into $\Lambda \overline{ K^0 }$.
The $N^0$ can be from the antidecuplet, 
from an octet \cite{octets_diakonov} or an 27-plet \cite{ellis}, while
the $\Xi^0$ I=1/2  from an octet.
In  \cite{ellis}  the  best estimate of 
m($N$) is  $\sim$ 1730 MeV.
The mass of the $N^0$ is expected in the approx. range 1650-1780 MeV 
\cite{arndt0312126,octets_diakonov,ellis}.
The $\Xi^0$ I=1/2 is expected to be near 1700 
 \cite{octets_diakonov} 
or it maybe degenerate with the $\Xi^0(1860)$ I=3/2  \cite{jaffe}.
The fact that 
we don't observe a peak above background near 1850 or 1860 MeV
 disfavours the latter possibility.
However the branching ratio of a possible $\Xi^0(1850-1860)$ I=1/2
to $\Lambda K^0_s$ may be small and we
need more statistics to observe it.
\noindent
The mass of the peak at 1734 MeV  is in very good agreement with the 
$N$ mass of $\sim$
 1730 MeV suggested by Arndt et al \cite{arndt0312126}. In this paper
a modified Partial Wave analysis allows to search for narrow
states and presents two candidate $N$ masses, 1680 and/or 1730 MeV
with width below 30 MeV.

\noindent
We don't yet observe a pronounced peak at 1734 or 1693 MeV
 in the anti-channel $\overline{\Lambda} K^0_s$. This is work in progress.
If pentaquarks would 
be primarily formed through quark coalescence rather than through
hadronic interactions in Au+Au collisions at RHIC,
the expected antipentaquark to pentaquark ratio is estimated as follows:
\noindent
$
\overline{ \Xi^0} / \Xi^0 \sim
        \frac
        { \overline{u} \overline{d} \overline{s} \overline{s} d }
        { u d s s \overline{d} }
        \sim
        \frac { \overline{q} } {q}  \cdot (\frac {\overline{s}} {s} ) ^2
		\sim 0.90
$
respectively
\noindent
$
\overline{ N^0_s} / N^0_s \sim
        \frac
        { \overline{u} \overline{d} \overline{s} \overline{d} s }
        { u d s d \overline{s} }
        \sim
        (\frac { \overline{q} } {q})^3
		\sim 0.73
$
while we used a ratio of $\overline{p}/p$ = 0.73 \cite{star_plb567_2003_167}
and assumed $\overline{s}/s$ =1.
These values are at production, while further reduction of the
ratios can follow from e.g. absorbtion of the decay products.
Therefore, the non-observation of an
antiparticle may favour the $N^0$ hypothesis.
\noindent
The possible peak we observe at 1693 $\pm$ 0.5 MeV
is a candidate for the state $\Xi(1690)$ with mass 1690 $\pm$ 10 MeV
and width below 30 MeV \cite{pdg}.
If so, we improve the PDG limit of the width from $\Gamma < $ 30 MeV at present, to
$\Gamma < $ 6 MeV.

\vspace*{-0.3cm}

\section{Conclusions}\label{concl}
 
\vspace*{-0.2cm}
\noindent
We present preliminary results of the STAR experiment
 at RHIC on a search for the 
$N^0$ and $\Xi^0$ I=1/2 pentaquarks in minimum bias
Au+Au collisions at 200 GeV through the decay channel
$\Lambda K^0_s$.
We observe a  peak 
at a mass of  1733.6 $\pm$ 0.5 (stat) $\pm$
5 MeV (syst) and width  consistent with  the experimental resolution of about
6 MeV within the errors, which gains in significance when restricting the upper
$\sim$
 10\% of $\sigma_{tot}$.
We obtain an estimate of the significance of
 $S/ \sqrt{ B} $ = $30.6/ \sqrt{ 35.4}$ = 5.1,  $S/ \sqrt{S+B} $ = 3.8
and $S/\sigma(S)$ = 3.04.
The best significance  of $S/ \sqrt{ B} $ = $19.4/ \sqrt{ 10.6}$= 5.93
is obtained for semiperipheral events.
Systematic studies suggest that this peak is not due to
 missidentifications or split tracks or to the decay of $\Delta(1232)$.
If this peak
corresponds to a real particle state it would be 
a candidate for the 
$N^0$ (octet, antidecuplet or 27-plet)
 and $\Xi^0$ I=1/2 (octet) pentaquark states.
We don't observe a peak near 1850 - 1860 MeV, disfavouring
the picture of degenerate octet and antidecuplet 
even though a low branching ratio to $\Lambda K^0_s$ may prevent
us from observing a peak.
The non observation of a possible antiparticle, while this is work in progress,
may favour the $N^0$ hypothesis and dominant production through
quark coalescence.
\noindent
The ambiguity
between $\Xi^0$ and $N^0$ 
can be resolved by searching for their isospin partners
$N^+ \rightarrow \Lambda K^+$ and
$\Xi^- \rightarrow \Lambda K^-$.
This is work in progress as well as searches for the 
$\Theta^+$ \cite{sevil_qm2004}.
\noindent
Furthermore, 
new data taken in our 2004 run will enhance the
statistics of minimum bias Au+Au events by a factor $\sim$ 10-15.

\vfill\eject
\end{document}